\documentclass[letter]{aa}

\usepackage{graphicx,lscape,rotating,amssymb}
\usepackage{txfonts}

\begin{document}
   \titlerunning{ISOCAM-CVF Spectroscopy of YSOs}
   \title{ISOCAM-CVF Spectroscopy of the Circumstellar Environment 
\\ of Young Stellar Objects\thanks{Based on observations with ISO, an ESA project with instruments funded by ESA Member States (especially the PI countries: France, Germany, the Netherlands and the United Kingdom) and with the participation of ISAS and NASA.}}

   \author{
	R.D.~Alexander
          \inst{1}\fnmsep \thanks{\emph{Present address:} Institute of Astronomy, Madingley Road, Cambridge, CB3 0HA, UK.}
        \and
        M.M.~Casali\inst{1}
	\and
	P.~Andr\'{e}\inst{2}
	\and
	P.~Persi\inst{3}
	\and
	C.~Eiroa\inst{4}
          }

   \offprints{Richard Alexander, 
	\email{rda@ast.cam.ac.uk}}

   \institute{
	UK Astronomy Technology Centre, Royal Observatory, Blackford Hill, Edinburgh, EH9 3HJ, UK
     	\and
	Service d'Astrophysique, CEA/DSM/DAPNIA, C.E.~Saclay, F-91191 Gif-sur-Yvette Cedex, France
	\and
	IASF, CNR Roma, Via Fosso del Cavaliere, 00133 Roma, Italy
	\and
	Dpto. F\'{\i}sica Te\'{o}rica, C-XI, Facultad de Ciencias, UAM, Cantoblanco, 28049 Madrid, Spain
             }

   \date{Received - / Accepted -}


   \abstract{
We present the results of a mid-infrared (5--16.5$\mu$m) imaging spectroscopy survey of Young Stellar Objects (YSOs) and their surrounding environment in four low-mass star formation regions: RCrA, $\rho$ Ophiuchi, Serpens and Chamaeleon I.  This survey was performed using ISOCAM and its Circular Variable Filters (CVF) and observed 42 YSO candidates: we were able to obtain complete 5--16.5$\mu$m spectra for 40 of these with a spectral resolving power of $\lambda/\Delta\lambda\simeq40$.  A number of spectral features were measured, most notably the 9.7$\mu$m silicate feature, the bending modes of both water and CO$_2$ ices at 6.0 and 15.2$\mu$m respectively and the well-known unidentified feature at 6.8$\mu$m.  The strength of the unidentified feature was observed to correlate very well with that of the water ice bending mode and far less strongly with the CO$_2$ ice bending mode.  This suggests, in a manner consistent with previous observations, that the carrier of the unidentified feature is a strongly polar ice.  Absorption profiles of the bending mode of CO$_2$ ice are observed to show a significant long wavelength wing, which suggests that a significant fraction of the CO$_2$ ice observed exists in a polar (H$_2$O-rich) phase.  The sources observed in RCrA, $\rho$ Oph and Serpens show similar spectral characteristics, whilst the sources observed in Cha I are somewhat anomalous, predominantly showing silicate emission and little or no absorption due to volatile ices.   However this is consistent with previous studies of this region of the Cha I cloud, which contains an unusual cluster of YSOs.  From comparisons of the strengths of the water ice and silicate bands we detect an apparent under-abundance of water ice towards the sources in $\rho$ Oph, relative to both RCrA and Serpens.  This may be indicative of differences in chemical composition between the different clouds, or may be due to evaporation.  Finally the CO$_2$:H$_2$O ice ratios observed towards the sources in $\rho$ Oph show significantly greater scatter than in the other regions, possibly due to varying local conditions around the YSOs in $\rho$ Oph.

   }

   \maketitle


\section{Introduction}\label{sec:intro}
The circumstellar environment of young stars is difficult to study, as in its early phases a young accreting and evolving star can be surrounded by sufficient gas and dust to cause a large extinction. However, at visible and near-IR wavelengths this continuum extinction is smooth and contains little information on the nature of the surrounding gas and dust.  The re-radiated energy is emitted mainly at wavelengths longer than 2$\mu$m.  Molecular vibrational-rotational gas-phase and solid-state transitions occur in the mid-IR, and a number of different molecules have been detected towards YSOs (see the review by van Dishoeck \& Hogerheijde \cite{vDH99}).  Spectroscopy of YSOs has also confirmed the presence of a number of molecular ices, such as ices of water, CO$_2$ and CH$_3$OH (eg.~Gibb et al.~\cite{gibb00}; Boogert et al.~\cite{boogert00}).  In addition we also see both absorption and emission from refractory solids such as silicates and PAHs, and infrared continuum emission.  This continuum emission results from the reprocessing of stellar radiation by the dense dust envelopes of the YSO, and this infrared excess is often used as a criterion for the identification of YSOs in photometric surveys. 

Unfortunately, studies from the ground are constrained to work within atmospheric windows. In some cases, such as CO$_2$, telluric absorption by the molecule at 15.2$\mu$m makes its study in astronomical objects impossible. Furthermore, the sensitivity of observations in the mid-IR is greatly reduced because telescope and atmospheric thermal emission greatly increase the photon shot noise.  Consequently, ground-based studies longward of K band are generally restricted to observing relatively bright objects.  The only solution to these problems is to move to an airborne or space platform, where even a modest sized telescope can outperform ground-based systems.

Mid-IR spectra of young stars at high resolution were successfully taken with the Infrared Space Observatory (ISO, see Kessler et al.~\cite{iso} for an overview of the ISO mission) by various investigators (eg.~W33A by Gibb et al.~\cite{gibb00}; Elias 29 by Boogert et al.~\cite{boogert00}). In an attempt to push mid-IR observations to even higher levels of sensitivity we made use of ISOCAM (Cesarsky et al.~\cite{isocam}) and the Circular Variable Filters (CVF) to image 42 YSO candidates in four well-known low-mass star formation regions: RCrA, $\rho$ Ophiuchi, Serpens and Chamaeleon I.  The high sensitivity and good sample size was achieved by using the low CVF resolution and by observing dense clusters of YSOs, so that ISOCAM's imaging capability allowed simultaneous spectra to be obtained. It should be noted that the low spectral resolving power of the CVF ($\lambda$/$\Delta\lambda\simeq40$) essentially prohibits the detection of gas-phase spectral features; this is a study of the molecular ices and dust that lie around these young stars.  However it is possible to draw a number of conclusions about the physics and chemistry around these YSOs, with information about their composition, thermal behaviour and evolution readily obtainable from such spectroscopic observations.


\section{Observations and Data Reduction}\label{sec:obsreduc}
\subsection{Observations}\label{sec:obs}
5--16.5$\mu$m imaging spectroscopy was performed on six target fields in four low-mass star formation regions using the ISOCAM-CVF, on various dates between April 1996 and February 1997.  This instrument made use of 2 variable width filters to take images at a number of different wavelengths, using a 32$\times$32 pixel gallium-doped silicon array.  Exposure times of 2.1s per CVF step were used.  Each was exposed repeatedly and the images were sequenced (in time) in decreasing order of wavelength, resulting in a total exposure time of $\sim$2500s per target region for each CVF set.  This procedure produced 2 datacubes for each target region. The first covered the wavelength range 5.0--9.4$\mu$m (CVF1) and the other covered 9.3--16.3$\mu$m (CVF2), with a mean spectral resolving power of $\lambda / \Delta\lambda \simeq 40$.  A brief summary of the characteristics of the observed regions is given below.

\subsubsection{RCrA}\label{sec:rcra}
The RCrA cloud is the name given to a molecular cloud of mass $\sim$120M$_{\sun}$ (Harju et al.~\cite{harju93}) around the variable star \object{R Coronae Australis}, at a distance of 130pc (Marraco \& Rydgren \cite{mr81}).  It is a region of low-mass star formation located approximately 18$\degr$ below the galactic plane (Olofsson et al.~\cite{olofsson99}), where it suffers little foreground obscuration.  The previous K-band studies of Taylor \& Storey (\cite{ts84}) and Wilking et al.~(\cite{wilking97}) and the sub-millimetre survey of Harju et al.~(\cite{harju93}) revealed a number of embedded objects and other objects classified as YSOs.  One field of view was observed in the \object{RCrA cloud}, using the 6$\arcsec$ pixel field of view (pfov).  It was centred on 19$^{\rm h}$01$^{\rm m}$45$\fs$1, $-$36$\degr$58$\arcmin$34$\arcsec$\footnote{Please note that all coordinates given are J2000.0.}, immediately to the south-west of \object{RCrA} itself.

\subsubsection{$\rho$ Ophiuchi}\label{sec:rhooph}
The $\rho$ Ophiuchi dark cloud complex is a complicated structure of several large molecular clouds near the bright star \object{$\rho$ Ophiuchi}.  Its relatively high galactic latitude ($b$=$17^{\circ}$) combined with its relative proximity ($d$=160pc, Whittet \cite{whittet74}) result in comparatively little foreground obscuration, making it an ideal region in which to study star formation.  Initially identified by Grasdalen et al.~(\cite{gss73}), who identified 41 sources in a K-band map, it has since been studied extensively both in the infrared (Elias \cite{elias78}; Greene \& Young \cite{gy92}; Strom et al.~\cite{sks95}; Abergel et al.~\cite{abergel96}; Bontemps et al.~\cite{bontemps01}) and at millimetre wavelengths (Loren et al.~\cite{lww90}; Andr\'{e} \& Montmerle \cite{am94}).  In particular, Loren et al.~(\cite{lww90}) showed that the cloud consists of a number of smaller sub-structures.  They identified 6 dense cores (labelled A-F) and observed significantly higher clustering of YSOs in three of these cores (A, B, and E/F).  Two fields were observed in $\rho$ Oph, both with the 6$\arcsec$ pfov: one centred on 16$^{\rm h}$26$^{\rm m}$21$\fs$4, $-$24$\degr$23$\arcmin$59$\arcsec$ (field $\rho$ Oph A) and one at 16$^{\rm h}$27$^{\rm m}$23$\fs$7, $-$24$\degr$40$\arcmin$39$\arcsec$ (field $\rho$ Oph E).  The positions correspond (approximately) to cores A and E/F above.

\subsubsection{Serpens}\label{sec:ser}
The \object{Serpens molecular cloud} was initially identified by Strom et al.~(\cite{svs76}).  It has since been very well surveyed, both in the infrared (Churchwell \& Koornneef \cite{ck86}; Gomez de Castro et al.~\cite{gel88}; Eiroa \& Casali \cite{ec92}; Giovannetti et al.~\cite{giov98}; Kaas \cite{kaas99}) and the sub-millimetre (Casali et al.~\cite{casali93}), with as many as 163 sources identified in the 6'$\times$5' K-band map of Eiroa \& Casali (\cite{ec92}).  The cloud itself extends over a region of $\ge$20'$\times\ge$15' in CO line emission (Loren et al.~\cite{loren79}) and is at a distance of $\simeq$260pc (Strai\v{z}ys et al.~\cite{scb96}).  
In addition to survey imaging, many individual objects in the cloud have been studied in detail: Eiroa \& Casali (\cite{ec89}) studied the multiple outflow source \object{SVS4}; \object{SVS2} has been shown to illuminate the \object{Serpens Reflection Nebula} (Gomez de Castro et al.~\cite{gel88}; Huard et al.~\cite{hwk97}); \object{SVS20} has been shown to be a close binary system  (Huard et al.~\cite{hwk97}).
Two fields were observed in Serpens.  Field Ser B, taken with the 6$\arcsec$ pfov, was of the region around \object{SVS2} and \object{SVS20} and was centred on 18$^{\rm h}$29$^{\rm m}$57$\fs$6, 1$\degr$12$\arcmin$42$\arcsec$.  Field Ser A was a higher resolution field (3$\arcsec$ pfov) of the multiple outflow source \object{SVS4}, centred on 18$^{\rm h}$29$^{\rm m}$57$\fs$2, 1$\degr$14$\arcmin$25$\arcsec$.

\subsubsection{Chamaeleon I}\label{sec:cha}
The \object{Chamaeleon dark cloud} complex is a complicated structure consisting of 3 large molecular clouds (designated Cha I,II,III by Hoffmeister \cite{hoffmeister63}) and a number of smaller clumps and globules. Again, the relative proximity ($d$=160pc, Whittet et al.~\cite{whittet97})  and high galactic latitude ($b\sim-$16$\degr$) of the clouds result in little foreground obscuration and make them ideal for star formation study.  The region has been well-surveyed at millimetre wavelengths (Mattila et al.~\cite{mlt89}; Henning at al.~\cite{henning93}) and more recently in the near- (Kenyon \& G\'{o}mez \cite{kg01}) and mid-infrared (Persi et al.~\cite{persi00}).  A single field was observed in the \object{Cha I cloud}, with the 6$\arcsec$ pfov, centred on 11$^{\rm h}$09$^{\rm m}$39$\fs$4, $-$76$\degr$35$\arcmin$2$\arcsec$.  The corresponds (approximately) to a previously observed dense core lying to the north of \object{HD97300} (Jones et al.~\cite{jones85}; Persi et al.~\cite{persi99}).

\subsection{Data Reduction}\label{sec:reduc}
Initial data reduction was performed using the auto-analysis package from the ISO data pipeline (OLP V10.0, Ott et al.~\cite{pipeline}; Blommaert et al.~\cite{handbook}).  This corrected the images for changes in the spectral response of the detector, as well as performing flatfielding and flux calibration.  It should be noted that column 24 on the array was dead: this was corrected for by means of a simple linear interpolation from the adjacent pixels.  Sources were identified ``by eye'' and everything representing a local maximum in intensity in more than 75\% of the frames was treated as a source.  For each source identified the flux in each frame was evaluated as the total flux contained within a 18$\arcsec$$\times$18$\arcsec$ (3$\times$3 pixel) aperture\footnote{A 15$\arcsec$$\times$15$\arcsec$ (5$\times$5 pixel) aperture was used for the 3$\arcsec$ pfov frames of field Ser A.} less the sky flux (evaluated as the mean flux per pixel over a designated sky region).  This produced a spectrum for each source from both filter sets.  These spectra were smoothed to match the resolution of the CVF and the 2 ``halves'' taken from the 2 filter sets were then joined together.  Unfortunately it was necessary to discard the data from positions 70--80 on CVF1 (8.94--9.43$\mu$m), as the array ``memory'' (a noted problem with ISO, see Blommaert et al.~\cite{handbook}) rendered these images useless.  Consequently the spectra contain a small gap in their wavelength coverage at this point, from 8.94--9.33$\mu$m.  The spectra were also corrected for a systematic under-estimation of flux at long wavelength, which arose from the use of a fixed-size aperture to measure the total flux from diffraction-limited images.

\subsection{Spectral Features}\label{sec:features}
\begin{figure}
	\resizebox{\hsize}{!}{
	\begin{turn}{270}
	\includegraphics{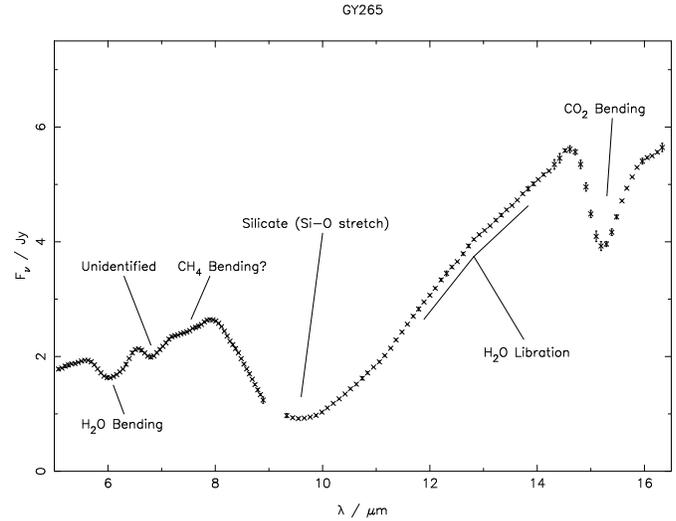}
	\end{turn}
	}
	\caption{A typical spectrum, with prominent spectral features annotated.}
	\label{fig:typ}
\end{figure}
An example of a typical spectrum is shown in Fig.\ref{fig:typ}.  The following spectral features were identified in the spectra:
\begin{itemize}
\item An absorption feature at 6$\mu$m attributed to the bending mode of H$_2$O ice (Gerakines et al.~\cite{gerakines95}).
\item An absorption feature at 6.8$\mu$m, the well-known unidentified feature (Schutte \cite{schutte97}).
\item A weak absorption feature at $\sim$7.6$\mu$m which was not resolved at the low spectral resolution of the CVF.  It is tentatively identified as the bending mode of CH$_4$ at 7.67$\mu$m (Hudgins et al.~\cite{hudgins93}; Boogert et al.~\cite{boogert96}; Boogert et al.~\cite{boogert98}) but due to the low resolution of the CVF and the weak nature of the observed feature this identification is somewhat uncertain.
\item A broad feature centred at $\sim$9.7$\mu$m, seen in both emission and absorption, identified as the Si--O stretching mode of ``astronomical silicate'' (Draine \& Lee \cite{dl84}; Simpson \cite{simpson91}; O'Donnell \cite{odonnell94}).
\item A broad absorption feature at $\sim$13$\mu$m identified as the libration (or hindered rotation) mode of H$_2$O ice (Gerakines at al.~\cite{gerakines95}).
\item An absorption feature at 15.2$\mu$m attributed to the bending
mode of CO$_2$ ice (Ehrenfreund et al.~\cite{ehren96}).
\end{itemize}

Other spectral features were observed in individual sources, such as the apparent 10.5$\mu$m emission feature in RCrA IRS5 (see Fig.\ref{fig:goodfits}) or the apparent 14$\mu$m absorption feature in HH100-IR (see Fig.\ref{fig:hh100ir}).  These rarer features have been noted in the results tables where they occur, but the goal of this study was to look for general trends rather than individual curiosities.  With this in mind, and also considering the large number of spectra obtained, it was decided to fit only the strongest and most commonly occuring features highlighted above.

\subsection{Fitting the Spectra}\label{sec:fitting}
As many of these spectral features overlap, independent determinations of their strengths were not possible, so all of the features in the spectra were fitted simultaneously.  This was achieved by fitting a spectrum of the form:
\begin{equation}
F_{\nu}(\lambda) = C(\lambda)\times\exp\left(-\sum_i s_i \alpha_i(\lambda)\right)
\end{equation}
where $C(\lambda)$ is the continuum, $s_i$ represents the absorbing column density of each species $i$ and $\alpha_i(\lambda)$ are the absorption coefficients for each species $i$.  This essentially treats the absorbing material as a cold absorbing screen.  In order to account for possible silicate emission, the depth of the silicate feature, $s_{\mathrm {Si}}$, was allowed to run negative.  Note however that this assumes that the same profile applies in both absorption and emission, so could result in errors if both are present.  The absorption coefficients for H$_2$O ice and CO$_2$ ice were taken from the Leiden Observatory Database of Interstellar Ices (Gerakines et al.~\cite{gerakines95}; Ehrenfreund et al.~\cite{ehren96}) and those coefficients evaluated at 10K were used, as thermal variations are negligible at the spectral resolution of the CVF.  It should be noted, however, that the depths of the 6$\mu$m and 13$\mu$m water ice bands did not correlate as expected, so they were fitted independently.  The unidentified feature at 6.8$\mu$m was fitted using an asymmetric Gaussian profile: a Gaussian profile with two different widths shortward and longward of the peak. The best-fitting profile for each 6.8$\mu$m feature was measured, and the widths were then fixed as the means of the widths measured in all of the observed profiles (FWHM=500nm).  The poorly resolved 7.6$\mu$m feature was fitted using a Gaussian profile with a width equal to that of the CVF resolution element at that wavelength.  A number of profiles exist in the literature for the silicate feature, obtained both theoretically and observationally, so the profile which provided the best fit to the observed data was used: profile 2 from Simpson (\cite{simpson91}).  However the profiles obtained by Simpson (\cite{simpson91}) from IRAS observations extend shortward only to $\simeq$7.5$\mu$m, and so shortward of this wavelength the optical constants for ``circumstellar silicates'' from Ossenkopf et al.~(\cite{ohm92}) were used.  It should be noted that the profiles from Simpson (\cite{simpson91}), Ossenkopf et al.~(\cite{ohm92}) and Draine (\cite{draine85}) are almost identical away from the 9.7$\mu$m peak: only the shape of the absorption band is dependent on the choice of optical constants.

The fitting procedure was iterative, with repeated fitting converging to a final profile.  The fitting algorithm took the form:
\begin{enumerate}
\item Fit initial continuum and divide out.  The continuum was fitted as a 2nd-order polynomial through points at 5.6, 7.9 and 16.1$\mu$m, so chosen as they best constrained the resulting fits.
\item Measure the strength of each spectral feature at a number of points around the peak.
\item Use these strengths and the optical constants for the different species to divide out the features from the spectra.
\item Fit a new continuum to this ``featureless'' spectrum.
\item Iterate this procedure until the fit converged: typically 10-15 iterations were sufficient.
\end{enumerate}

\begin{figure}
	\resizebox{\hsize}{!}{
	\includegraphics{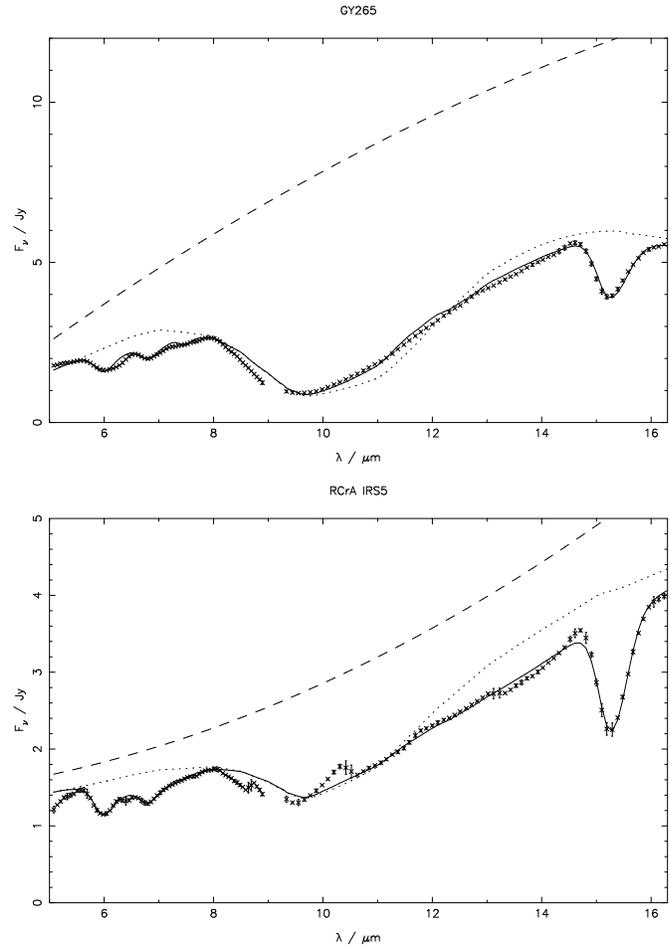}
	}
	\caption{Examples of successful fits to GY265 in $\rho$ Oph (upper) and RCrA IRS5 (lower): fitted spectra are shown as solid lines and continua as dashed lines.  The dotted lines are the ``continuum plus silicate'' fits given by $C(\lambda)\times\exp\left(- s_\mathrm{Si} \alpha_\mathrm{Si}(\lambda)\right)$.}
	\label{fig:goodfits}
\end{figure}

\noindent It should be noted that this allowed the continuum to drift away from the observed spectrum in some cases, due to absorption across the entire observed wavelength range.  This is because the normalised silicate opacity over the range 5-7.5$\mu$m is approximately constant at $\simeq$0.15, and towards the long wavelength end of the spectra the 18$\mu$m bending feature begins to ``cut in''.  (The normalised opacity at 16$\mu$m is approximately 0.3.)  The effect of this can be seen in Fig.\ref{fig:goodfits}: although the continua are seen to deviate significantly from the observed spectra, appearing to diverge from the observed spectra in some cases, this is primarily due to silicate absorption across the entire wavelength range.  With only the silicate features included, these ``continua plus silicate'' fits are seen to follow the observed spectra well.  As a result, any possible systematic errors arising from the manner of the continuum fit will only affect the evaluation of the silicate optical depth; the measured strengths of the narrower features are not affected by the choice of continuum.

On inspection of the residuals from this fitting procedure a further feature was observed: a broad feature centred on $\simeq$11.2$\mu$m, seen in both emission and absorption.  This has a number of possible identifications, such as crystalline silicates (Bregman et al.~\cite{bregman87}; Campins \& Ryan \cite{cr89}), the Unidentified (PAH) Band at 11.3$\mu$m (Boulanger et al.~\cite{boulanger96}), or a shoulder on the silicate profile.  Given the variation in possible identifications, it was decided to fit this feature independently using an asymmetric Gaussian profile, once again taking the widths to be the mean of those observed (FWHM=1700nm).  This feature was then included in the algorithm and the fitting procedure repeated: examples of successful fits are shown in Fig.\ref{fig:goodfits}.  Another similar residual around 8.5$\mu$m was also observed, but the properties and strength of this were greatly affected by the choice of continuum point around 8$\mu$m.  Whereas the 11$\mu$m feature was robust against small changes in the continuum this apparent 8.5$\mu$m feature was not, so it was not included in the fitting procedure.

The 15.2$\mu$m CO$_2$ ice feature is an interesting one and has been used in the past as a diagnostic of the ice environment.  Gerakines at al.~(\cite{gerakines99}) found that two distinct phases of CO$_2$ ice (polar and non-polar) show significantly different absorption profiles.  The non-polar phase is characterised by a double-peaked profile, arising from the two-fold degeneracy of the bending mode, whereas the polar (H$_2$O-rich) phase shows a significant long wavelength wing.  The spectral resolution of the CVF is too low to resolve the double-peak structure, but the CO$_2$ profiles were observed to vary noticeably from source to source.  As a result it was decided to let the CO$_2$ profile vary, so it was fitted using an asymmetric Gaussian profile with the widths as variable parameters.

There are two forms of error associated with this fitting procedure.  Firstly there are the ordinary statistical errors associated with the fits, evaluated by combining the statistical errors on the data points used to fit the features and the intrinsic uncertainties in the individual feature fits.  These are typically around 5--10\% for the stronger features (silicate, H$_2$O, CO$_2$) and around 15--20\% for the less well constrained features, although in the case of some poor S/N spectra they are much larger.  In addition to this, there are systematic errors associated with poor fits.  These are usually due to the presence of ``extra'' features in the spectra, or because of poorly fitting feature profiles, and are less easy to quantify.  Consequently, the errors reported in the results tables are the statistical errors only; where fits are flagged as poor additional systematic errors also apply. 

\subsection{Dealing with poor fits}\label{sec:badfits}
\begin{figure}
	\resizebox{\hsize}{!}{
	\includegraphics{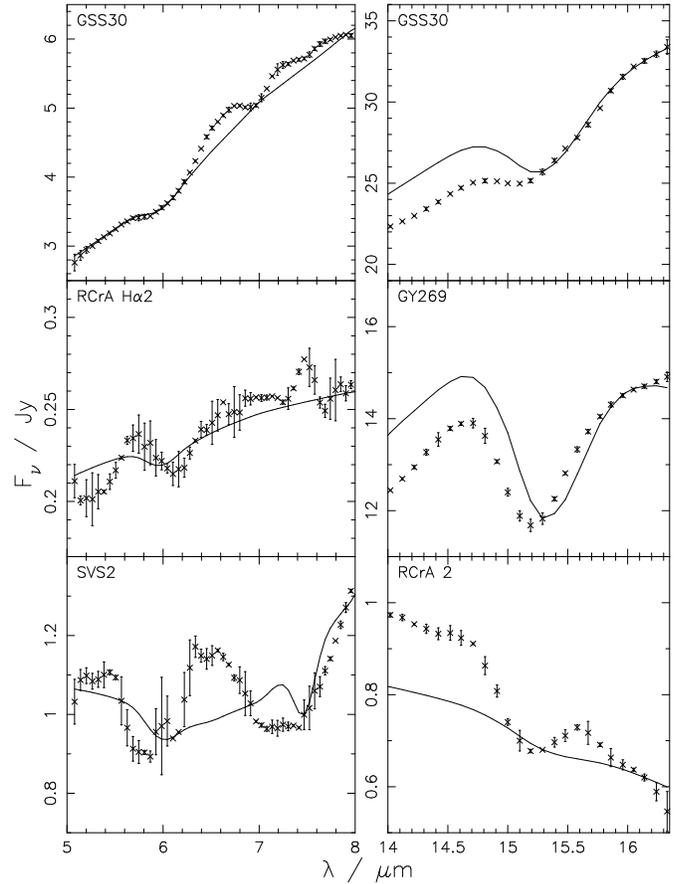}
	}
	\caption{Examples of poor fits (solid lines) from 5--8$\mu$m and around the CO$_2$ ice feature.}
	\label{fig:badfits}
\end{figure}
Generally the procedure described above proved successful, but in some cases it led to poor fitting of certain features.  This was primarily due to the manner in which the continuum was fitted, as shown in the examples in Fig.\ref{fig:badfits}.  The continuum was fitted as a 2nd-order polynomial through three sections on the spectrum expected to show the least absorption, as described above.  However, constraining the continuum through these three points led in some cases to poor fitting elsewhere, particularly at the short end of the CO$_2$ ice band and immediately shortward of the silicate feature.

In such cases, the narrower features at the short end of the spectrum (6.0, 6.8 and 7.6$\mu$m) were fitted in exactly the same way as before, only this time using a linear continuum through the points at 5.6 and 7.9$\mu$m.  As absorption due to the silicates is approximately constant over this wavelength range this removed complications due to the quadratic continuum fit.  When tested on spectra which were fitted well, such as those in Fig.\ref{fig:goodfits}, this produced results which were the same as before, within the fitting errors.  Also, the manner in which the continuum was fitted often led to the CO$_2$ feature fitting poorly, as seen in Fig.\ref{fig:badfits}. As a result it was decided to measure the CO$_2$ absorption in all the sources by fitting a linear continuum from 14.5--16.1$\mu$m, as the silicate absorption does not vary much over this range and the CO$_2$ is not blended with any other spectral feature.  This procedure was only applied to these narrower features, however, as the broad silicate, 11 and 13$\mu$m bands were deemed to be too heavily blended to deconvolve them meaningfully over a short section of the spectra.  In addition, all features fitted in this manner are flagged as such in the results tables.

\subsection{Spectral Index}\label{sec:index}
In order to take a standard measurement of the SED, the spectral index $\alpha$ was evaluated as a traditional IR spectral index (eg.~Kenyon \& Hartmann \cite{kh95}):
\begin{equation}
\alpha = \frac{\log \lambda_2 F_{\lambda_2} - \log \lambda_1 F_{\lambda_1}}{\log \lambda_2 - \log \lambda_1}
\end{equation}
This sets $\alpha=0$ as the index of a flat $\nu F_{\nu}$ and $\alpha=-3$ as the index of a Rayleigh-Jeans spectrum.  Equivalently, class I YSOs have $\alpha>0$, class II have $-3<\alpha<0$ and class III $\alpha \sim -3$, using the conventional definition of spectral class (Lada \cite{lada87}), although more recent studies (eg.~Andr\'{e} \& Montmerle \cite{am94}; Greene et al.~\cite{greene94}) have found that the class II/III border lies at a higher value ($\alpha \simeq -1.5$).  Where previous K-band data exist in the literature the wavelengths were taken as $\lambda_2=14.0$$\mu$m (our data) and $\lambda_1=2.2$$\mu$m (K-band data from references in Table \ref{tab:sources}) and this was applied to both the observed spectrum and the calculated continuum for each source.  (Where no K-band data exist, for sources not previously observed or with uncertain identifications, $\lambda_1=8.0$$\mu$m was used.)  This second (``continuum'') measurement of the spectral index is important, as many of the sources suffer significant absorption (including foreground absorption) and a straight application to the observed spectrum could result in sources being mis-classified in cases where deep absorption is present.  In cases where the flux suffers little extinction the two give the same results, but where deep absorption is present the continuum spectral index derived is greater than that obtained from the observed spectrum.  In such cases the continuum spectral index provides a truer measurement of the shape of the underlying continuum.


\section{Results}\label{sec:results}
\subsection{Detected Sources}\label{sec:detect}
A total of 43 sources were detected: a complete list is presented in Table \ref{tab:sources}.  The 6$\arcsec$ pfov used here results in coordinates which are accurate only to $\pm3$--$4\arcsec$ at best, so where sources have been identified with previous studies the coordinates presented are those from previous observations, where higher spatial resolutions or better S/N have resulted in more accurate positions.  The coordinates evaluated here are only presented for sources which represent new or uncertain detections.  In the 3 cases where sources have been identified as unresolved doubles the coordinates presented are those of the brighter source.  Of the 43 sources detected 6 were not identified with detections from previous surveys (see Sect.~\ref{sec:newsources}) and 2 (Ser A 7 and Ser B 11) were determined to be different images of the same source, as the two Serpens fields overlap slightly.  Consequently we were able to obtain complete 5.0--16.3$\mu$m spectra for a total of 40 sources, along with a further 2 partial spectra\footnote{These partial spectra were the result of various observational problems, such as sources lying near to the edge of the frame or lying on the bad column.}: a complete atlas of these 42 spectra is presented in Appendix \ref{sec:app}.

\begin{table*}
\begin{minipage}{\linewidth}
\caption{Detected sources: references indicate origin of specified coordinates.}\label{tab:sources}

\begin{tabular}{lcclcl}
\hline\hline
Source & $\alpha_{2000} \; (h\;m\;s)$ & $\delta_{2000}
\;(\degr\;\arcmin\;\arcsec)$ & Previous ID$^{\mathrm{a}}$ &
Reference$^{\mathrm{a}}$ & Notes\\\hline
RCrA 1 & 19 01 41.5 & $-$36 58 29 & \object{IRS2, TS13.1} & 2 & \\
RCrA 2$^{\mathrm{b}}$ & 19 01 41.5 & $-$36 58 59 & & 17\\
RCrA 3 & 19 01 41.7 & $-$36 59 54 & \object{H$\alpha$2} & 1 & \\
RCrA 4 & 19 01 48.0 & $-$36 57 19 & \object{IRS5, TS2.4} & 2 & \\
RCrA 5 & 19 01 50.7 & $-$36 58 07 & \object{HH 100-IR, IRS1, TS2.6} & 2 & \\
RCrA 6$^{\mathrm{b}}$ & 19 01 50.9 & $-$36 57 37 & & 17 & Very close to RCrA\\
RCrA 7$^{\mathrm{b}}$ & 19 01 52.4 & $-$36 57 37 & & 17 & Very close to RCrA\\
\hline
$\rho$ Oph A 1 & 16 26 17.3 & $-$24 23 49 & \object{SKS9, ISO-Oph 21}  & 6\\
$\rho$ Oph A 2 & 16 26 18.8 & $-$24 24 21 & \object{SKS11, ISO-Oph 26} ??? & 17 & Uncertain ID\\
$\rho$ Oph A 3 & 16 26 21.5 & $-$24 23 07 & \object{GSS30, GY6, SKS12, ISO-Oph 29} & 6\\
$\rho$ Oph A 4 & 16 26 23.7 & $-$24 24 40 & \object{GY21, SKS16, ISO-Oph 37} & 6 & Falls on bad column\\
\hline
$\rho$ Oph E 1 & 16 27 21.6 & $-$24 41 43 & \object{GY252, SKS36, IKT43, ISO-Oph 132} & 6\\
$\rho$ Oph E 2 & 16 27 24.8 & $-$24 41 02 & \object{ISO-Oph 137} & 6 \\
$\rho$ Oph E 3 & 16 27 26.9 & $-$24 39 23 & \object{GY262, IKT53, ISO-Oph 140} & 6 \\
$\rho$ Oph E 4$^{\mathrm{b}}$ & 16 27 27.1 & $-$24 41 31 & & 17 & Very faint \\
$\rho$ Oph E 5 & 16 27 27.2 & $-$24 40 49 & \object{GY265, IKT54, ISO-Oph 141} & 6 \\
$\rho$ Oph E 6 & 16 27 28.2 & $-$24 39 32 & \object{GY269, IKT57, ISO-Oph 143} & 6 \\
\hline
Ser A 1 & 18 29 56.6 & +1 12 56 & \object{SVS4/2, GEL5} & 11 & Very faint \\
Ser A 2 & 18 29 56.7 & +1 12 35 & \object{SVS4/3} & 11 & Very faint \\
Ser A 3 & 18 29 57.5 & +1 12 57 & \object{SVS4/5} {\bf and}  & 11 & Unresolved double\\
       &            &          & \object{SVS4/6, GEL9}\\
Ser A 4 & 18 29 57.9 & +1 12 25 & \object{SVS4/7} & 11 & Very faint \\
Ser A 5 & 18 29 57.9 & +1 12 34 & \object{SVS4/8} & 11 & \\
Ser A 6 & 18 29 57.9 & +1 12 43 & \object{SVS4/9} {\bf and}  & 11 & Unresolved double\\
        &            &          & \object{SVS4/10, GEL11} \\
Ser A 7$^{\mathrm{c}}$ & 18 30 00.1 & +1 13 03 & \object{GCNM130} & 13 & Off frame 7.0-9.3$\mu$m\\
\hline
Ser B 1 & 18 29 52.9 & +1 14 55 & \object{GCNM53} & 13 \\
Ser B 2 & 18 29 55.7 & +1 14 31 & \object{CK9, GEL4, EC74, GCNM76} & 13\\
Ser B 3 & 18 29 56.8 & +1 14 46 & \object{SVS2, CK3, GEL6, EC82, GCNM87} & 13 & Unresolved double\\
	&	     &          & {\bf and} \object{GEL8, EC86, GCNM93} \\
Ser B 4$^{\mathrm{b}}$ & 18 29 57.2 & +1 13 28 & & 17 & Peak in nebulosity?\\
Ser B 5 & 18 29 57.6 & +1 15 31 & \object{GCNM100} & 13 \\
Ser B 6 & 18 29 57.7 & +1 14 05 & \object{SVS20, CK1, GEL10, EC90, GCNM98} & 13 & Double Object \\
Ser B 7$^{\mathrm{b}}$ & 18 29 58.1 & +1 13 24 & & 17 & Peak in nebulosity?\\
Ser B 8 & 18 29 58.2 & +1 15 21 & \object{CK4, GEL12, EC97, GCNM106} & 13 & Falls on bad column\\
Ser B 9 & 18 29 58.7 & +1 14 26 & \object{EC103, GCNM112} & 13 \\
Ser B 10 & 18 29 59.2 & +1 14 08 & \object{CK8, GEL13, EC105, GCNM119} & 13 & \\
Ser B 11$^{\mathrm{c}}$ & 18 30 00.1 & +1 13 03 & \object{GCNM130} & 13\\
Ser B 12 & 18 30 00.5 & +1 15 20 & \object{CK2, EC118, GCNM136} & 13 & Peaks on bad column\\
Ser B 13 & 18 30 02.1 & +1 13 59 & \object{CK7, EC125, GCNM154} & 13 & Very faint \\
\hline
Cha I 1 & 11 09 23.3 & $-$76 34 35 & \object{C1-6, ISO-ChaI 189, KG82} & 15 & Bad column in CVF1\\
Cha I 2 & 11 09 29.2 & $-$76 33 30 & \object{ISO-ChaI 192, KG87} & 15 & Out of frame in CVF1\\
Cha I 3 & 11 09 42.6 & $-$76 35 01 & \object{C1-25, ISO-ChaI 199, KG93} & 15\\
Cha I 4 & 11 09 46.9 & $-$76 34 49 & \object{C1-24, ISO-ChaI 204, KG97} & 15\\
Cha I 5 & 11 09 54.1 & $-$76 34 26 & \object{C1-5, ISO-ChaI 223, KG109} & 15\\
Cha I 6 & 11 10 00.8 & $-$76 34 59 & \object{WW Cha, C1-7, ISO-ChaI 231, KG116} & 15\\
\hline

\end{tabular}

\begin{list}{}{}
\item[$^{\mathrm{a}}$] Names and references from: 1) Marraco \& Rydgren (\cite{mr81}) (H$\alpha$); 2) Taylor \& Storey (\cite{ts84}) (TS, IRS); 3) Grasdalen et al.~(\cite{gss73}) (GSS); 4) Greene \& Young (\cite{gy92}) (GY); 5) Strom et al.~(\cite{sks95}) (SKS); 6) Bontemps et al.~(\cite{bontemps01}) (ISO-Oph); 7) Imanishi et al.~(\cite{ikt01}) (IKT); 8) Strom et al.~(\cite{svs76}) (SVS); 9) Churchwell \& Koornneef (\cite{ck86}) (CK); 10) Gomez de Castro et al.~(\cite{gel88}) (GEL); 11) Eiroa \& Casali (\cite{ec89}) (SVS4/?); 12) Eiroa \& Casali (\cite{ec92}) (EC); 13) Giovannetti et al.~(\cite{giov98}) (GCNM); 14) Hyland et al.~(\cite{hjm82}) / Jones et al.~(\cite{jones85}) (C1-?); 15) Persi et al.~(\cite{persi00}) (ISO-ChaI); 16) Kenyon \& G\'{o}mez (\cite{kg01}) (KG); 17) this paper.
\item[$^{\mathrm{b}}$] See Sect.~\ref{sec:newsources} for a discussion of the sources not previously identified.
\item[$^{\mathrm{c}}$] These two sources from the two different Serpens fields were determined to be different images of the same source.
\end{list}

\end{minipage}
\end{table*}

\subsubsection{Ghost Images and New Detections}\label{sec:newsources}
As noted by Blommaert et al.~(\cite{handbook}) ghost images and stray-light are a significant problem when using the CVF, so care must be taken when identifying sources.  Ghost images arise in two ways and are found either around the true image or directly opposite it (relative to the optical axis).  Potential ghost images were identified by looking at the spectra of the objects, their flux densities and their positions relative to the instrument optics.  In this manner 7 objects initially identified as sources were reclassified as ghost images (4 in $\rho$ Oph E, 2 in Cha I and 1 in RCrA).  This leaves 6 ``new'' detections in our survey, which we now address:
\begin{list}{}{}
\item[{\bf RCrA:}] 3 sources in \object{RCrA} are identified as new detections: sources RCrA 2, 6 and 7.  Source RCrA 2 lies close to RCrA 1, separated by $\sim$30$\arcsec$, but shows sufficiently different spectral characteristics that it cannot be a ghost of this source.  It is therefore considered to be a real source, probably a YSO.  Both RCrA 6 and 7 lie in a region of nebulosity between the bright sources \object{HH100-IR} (RCrA 5) and \object{RCrA} (which lies just outside the image frame).  These are not considered to be ghost images, but may well represent peaks in the nebulous emission rather than true YSOs.  It should also be noted that this region of the RCrA cloud was not included in the ISOCAM survey of Olofsson et al.~(\cite{olofsson99}), due to detector saturation, so it is not unreasonable to expect new detections here.
\item[{\bf $\rho$ Oph E:}] A large number of ghost images were found here and after they were removed 1 unidentified source remained: $\rho$ Oph E 4.  However, this is the faintest source in the survey, with a flux rising to just 0.17Jy at 16$\mu$m, so is probably a peak in the nebulosity rather than an ``new'' YSO.
\item[{\bf Ser B:}]  2 sources in Ser B were not identified with previous observations: Ser B 4 and 7.  Both these sources lie near to \object{SVS20} and so could be ring-like ghost images associated with it.  However they also lie in a previously identified region of nebulous emission (Kaas \cite{kaas99}) and so may instead be peaks in this nebulosity. It is impossible to resolve this ambiguity completely, but in either case these 2 sources are probably not ``new'' YSOs.
\end{list}

\subsubsection{Background Stars}\label{sec:bgstars}
Of the 36 sources identified with previous studies 3 have previously been identified as stars which are not associated with the star-forming clouds: \object{CK2} (Ser B 12) by Chiar et al.~(\cite{chiar94}) and Casali \& Eiroa (\cite{ce96}); Ser B 1 and Ser B 11 / Ser A 7 by Giovannetti et al.~(\cite{giov98}).

\object{CK2} is the only one of these 3 sources which has been identified as a background field star (probably a background supergiant, Casali \& Eiroa \cite{ce96}).  Unfortunately it fell directly on the bad column in the array here and so no useful spectroscopic data regarding \object{CK2} was obtained.  Both of the other two sources in Serpens show extremely deep absorption features due to ices and silicates and both have very red SEDs.  Taking the K-band magnitudes of Giovannetti et al.~(\cite{giov98}) and assuming the SEDs of these stars to be Rayleigh-Jeans spectra ($F_{\nu}\propto \lambda^{-2}$) results in predicted mid-IR flux densities of $\sim$0.1mJy for both sources, approximately 1000 times less than what is observed.  Consequently it seems unlikely that these are background sources, so we interpret them to be deeply embedded objects: with this interpretation these 2 sources are 2 of the 3 most deeply embedded objects in this survey.

\subsection{Spectral Fitting}\label{sec:fitres}
The spectra of the 43 sources identified in Table \ref{tab:sources} were fitted using the method described above: the results of this procedure are presented in Table \ref{tab:res1} and the fits and continua obtained are presented in Appendix \ref{sec:app}.  The fitting procedure is both robust and unambiguous, converging to a good fit in most cases, but a few weaknesses exist.  Firstly, as discussed in Sect.~\ref{sec:features}, the procedure only measures the 7 spectral features observed in almost all of the sources and consequently does not fit some rarer features, such as an apparent 14$\mu$m absorption band observed in source $\rho$ Oph A 2.  Further, the profile fitted to each feature (except the CO$_2$) is not allowed to vary from source to source.  While this is broadly valid, a few exceptions led to some poor fits, such as the broadened 6.0 and 6.8$\mu$m bands in \object{SVS2} (see Fig.\ref{fig:badfits}), or a handful of sources which show broadened silicate features.  However, given the low spectral resolution of the CVF and the variation in the spectra over the large number of sources 




\begin{landscape}
\begin{table}

\caption{Fitting results: errors are given in parentheses (ie.~7.84(16) $\leftrightarrow$ 7.84$\pm$0.16); source names correspond to those in Table \ref{tab:sources}.}\label{tab:res1}

\begin{tabular}{cccccccccccccl}\hline\hline

& 8$\mu$m & 8$\mu$m & & &\multicolumn{5}{c}{Equivalent Line Widths (nm)} & Source & Cont. & \\
Source & flux & continuum & $\tau_{\mathrm {Si}}$$^{{\mathrm a}}$ &
$\tau_{11 \mu \mathrm {m}}$ & 6$\mu$m & 6.8$\mu$m & 7.6$\mu$m &
13$\mu$m & 15.2$\mu$m & $\alpha$ & $\alpha$ & Group$^{{\mathrm b}}$ & Notes\\
& (Jy) & flux (Jy) & & & H$_2$O & & CH$_4$?? & H$_2$O & CO$_2$ & & & \\\hline
RCrA 1 & 7.84(16) & 7.79(16) & $-$0.02(1) & $-$0.06(1)& 178(8)$^{{\mathrm c}}$ & 17(1)$^{{\mathrm c}}$ & 17(3)$^{{\mathrm c}}$ & 247(25) & 168(7) & 0.56 & 0.58 & b & Strong 11$\mu$m emission \\

2 & 0.95(2) & 1.24(3) & 0.72(3) & $-$0.27(4) & 203(8) & 57(4) & 17(3) & 0(0) & 83(4) & $-$0.96$^{{\mathrm f}}$ & $-$1.35$^{{\mathrm f}}$ & a & Silicate too wide at long  \\
&&&&&&&&&&&&& $\lambda$, probable 13$\mu$m error \\ 

3 & 0.27(2) & 0.28(2) & 0.22(1) & $-$0.16(2) & 147(15)$^{{\mathrm c}}$ & 14(3)$^{{\mathrm c}}$ & 0(0)$^{{\mathrm c}}$ & 718(76) & 0(0) & $-$1.11 & $-$0.96 & b & $\tau_\mathrm{Si}$ probably too small, \\
&&&&&&&&&&&&& possible absorption bands\\
&&&&&&&&&&&&& at 13$\mu$m, 14.25$\mu$m\\

4 & 1.74(110) & 2.28(144) & 0.71(3) & $-$0.07(1) & 271(10) & 81(6) &
6(1) & 586(59) & 253(11) & 1.23 & 1.43 & a & Possible 10.5$\mu$m emission \\

5 & 22.46(44) & 36.91(74) & 1.35(5) & $-$0.23(3)& 206(8) & 67(4) &
14(2) & 1535(155) & 129(5) & 1.30 & 1.79 & a & \\

6 & 0.94(3) & 1.36(4) & 0.95(4) & $-$0.15(2) & 204(7) & 61(4) & 11(2)
& 73(7) & 97(4) & $-$0.40$^{{\mathrm f}}$ & $-$0.42$^{{\mathrm f}}$ & a & Under-estimates 13$\mu$m \\

7 & 1.20(27) & 1.62(37) & 0.74(3) & $-$0.08(1) & 191(9) & 86(6) & 18(3) & 0(0) & 95(4) & $-$0.18$^{{\mathrm f}}$ & $-$0.37$^{{\mathrm f}}$ & a & Possible 10.5$\mu$m emission, \\
&&&&&&&&&&&&& under-estimates 13$\mu$m
\\\hline 
$\rho$ Oph A 1 & 0.42(2) & 0.85(4) & 2.11(11) & $-$0.27(4) & 27(3) & 72(7) & 19(3) & 927(94) & 57(3) & $-$0.70 & $-$0.20 & a & Poor 5--8$\mu$m fit \\

2 & 0.21(2) & 0.54(6) & 2.68(19) & $-$0.70(10) & 207(14) & 253(53) & 79(9) & 2611(319) & 0(0) & 0.34 & 1.36 & a & Definite 14$\mu$m band \\

3 & 6.03(7) & 8.99(10) & 1.05(4) & $-$0.20(3) & 107(6)$^{{\mathrm c}}$ & 0(0)$^{{\mathrm c}}$?? & 0(0)$^{{\mathrm c}}$ & 10(3) & 87(3) & 1.61 & 1.84 & a & Possible 7.1$\mu$m band\\ 

4$^{{\mathrm e}}$ & 2.33(3) & 2.74(4) & 0.49(2) & $-$0.16(2) & 0(0) & 25(12) & 27(4)  & 687(69) & 23(1) & 0.95 & 1.15 & b & Silicate fit too wide \\
\hline
$\rho$ Oph E 1 & 2.88(3) & 3.36(4) & 0.52(2) & $-$0.21(3) & 140(5) & 60(4) & 27(4) & 1206(121) & 98(4) & 0.15 & 0.43 & b & Silicate too wide \\

2 & 0.23(2) & 0.56(5) & 2.55(12) & $-$0.36(5) & 255(16) & 93(7) & 53(8) & 1881(196) & 324(20) & $-$0.26$^{{\mathrm f}}$ & 0.51$^{{\mathrm f}}$ & a & Deep 7.6$\mu$m band \\

3 & 0.34(38) & 0.65(72) & 1.84(12) & $-$0.27(4) & 131(10) & 134(13) & 15(2) & 1536(160) & 155(7) & $-$0.06 & 0.47 & a & Very deep 6.8$\mu$m band \\

4 & 0.10(2) & 0.17(4) & 1.42(8) & $-$0.23(4) & 291(41)$^{{\mathrm c}}$ & 0(0)$^{{\mathrm c}}$ & 29(5)$^{{\mathrm c}}$ & 1561(279) & 501(49) & $-$0.55$^{{\mathrm f}}$ & $-$0.01$^{{\mathrm f}}$ & a & Very noisy: probably only \\
&&&&&&&&&&&&& class, CO$_2$, $\tau_\mathrm{Si}$ good data \\

5 & 2.62(4) & 5.89(9) & 2.22(8) & $-$0.29(4) & 218(14) & 105(7) & 20(3) & 339(34) & 207(8) & 1.17 & 1.59 & a & \\ 

6 & 4.60(25) & 16.10(89) & 3.31(13) & $-$0.52(7) & 306(12) & 117(7) & 0(0) & 193(20) & 112(4) & 1.54 & 2.17 & a & \\ 

\hline
Ser A 1 & 0.06(1) & 0.15(3) & 2.14(31) & $-$0.60(9) & 845(849)$^{{\mathrm c}}$ & 356(161)$^{{\mathrm c}}$ & 0(0)$^{{\mathrm c}}$ & 2386(311) & 298(33)? & $-$0.24 & 0.47 & a & Noisy and faint, possible \\
&&&&&&&&&&&&& 12, 14$\mu$m bands \\

2 & 0.06(1) & 0.17(3) & 2.39(61) & $-$0.76(10) & 965(878)$^{{\mathrm c}}$ & 298(130)$^{{\mathrm c}}$ & 0(0)$^{{\mathrm c}}$ & 2679(437) & 0(0) & 0.43 & 1.35 & a & Noisy, possible absorption \\
&&&&&&&&&&&&& bands at 11--16$\mu$m \\

3 & 0.64(2) & 1.24(4) & 1.70(7) & $-$0.32(4) & 523(20) & 165(11) & 19(3) & 1633(163) & 285(13) & 1.35 & 1.91 & a & Possible narrow absorption \\
&&&&&&&&&&&&& band at 9.7$\mu$m - CH$_3$OH? \\

4 & 0.07(1) & 0.15(2) & 2.14(34) & $-$0.46(7) & 847(688)$^{{\mathrm c}}$ & 271(69)$^{{\mathrm c}}$ & 0(0)$^{{\mathrm c}}$ & 2426(552) & 269(51)? & 0.42 & 1.17 & a & Noisy, possible absorption \\
&&&&&&&&&&&&& bands at 11--16$\mu$m \\

5 & 0.17(1) & 0.30(2) & 1.42(12) & $-$0.37(5) & 435(26) & 165(11) & 8(1) & 1538(177) & 186(11) & 0.25 & 0.71 & a & Possible 12, 14$\mu$m bands \\

6 & 0.46(1) & 0.45(1) & $-$0.18(1) & $-$0.22(3) & 272(17) & 97(6) & 0(0) & 803(80) & 223(9) & 0.44 & 0.50 & b & Strong 11$\mu$m emission \\

7 & - & - & - & - & 954(312)$^{{\mathrm d}}$ & - & - & - & 504(59) & - & - & a & \\
\hline
\end{tabular}

\begin{list}{}{}
\item[$^{{\mathrm a}}$] Values of $\tau_{\mathrm {Si}}$ between $-$0.8 and 0.6 are considered to be questionable, due to the presence of significant silicate emission and absorption (see discussion in Sect.~\ref{sec:si_prof}).
\item[$^{{\mathrm b}}$] Classification scheme proposed in Sect.~\ref{sec:si_prof}.
\item[$^{{\mathrm c}}$] Short wavelength features fitted using linear
continuum, fitted from 5.6--7.9$\mu$m, in cases where the automatic fitting
procedure produced errors (see Sect.~\ref{sec:badfits}).
\item[$^{{\mathrm d}}$] Feature fitted manually as spectrum was incomplete.
\item[$^{{\mathrm e}}$] Source fell on bad column on array: data from these sources is not considered to be especially accurate.
\item[$^{{\mathrm f}}$] No K-band data: $\alpha$ evaluated using $\lambda_1 = 8.0$$\mu$m.
\end{list}

\end{table}
\end{landscape}
\begin{landscape}
\begin{table}
\begin{tabular}{cccccccccccccl}\hline\hline

& 8$\mu$m & 8$\mu$m & & &\multicolumn{5}{c}{Equivalent Line Widths (nm)} & Source & Cont. & \\
Source & flux & continuum & $\tau_{\mathrm {Si}}$$^{{\mathrm a}}$ & $\tau_{11 \mu \mathrm {m}}$ & 6$\mu$m &
6.8$\mu$m & 7.6$\mu$m & 13$\mu$m & 15.2$\mu$m & $\alpha$ & $\alpha$
& Group$^{{\mathrm b}}$ & Notes\\
& (Jy) & flux (Jy) & & & H$_2$O & & CH$_4$?? & H$_2$O & CO$_2$ & & & \\\hline
Ser B 1 & 0.19(1) & 0.57(3) & 2.76(16) & $-$0.39(5) & 703(44) &
215(31) & 0(0) & 962(99) & 267(14) & 2.02 & 2.63 & a & Very deep absorption features \\ 

2 & 0.19(1) & 0.16(1) & $-$0.41(2) & $-$0.23(3) & 91(5)$^{{\mathrm c}}$ & 39(3)$^{{\mathrm c}}$ & 25(4)$^{{\mathrm c}}$ & 0(0) & 67(10) & 0.14 & 0.04 & b & Large discontinuity at join, \\ 
&&&&&&&&&&&&& poor fit from 8--12$\mu$m \\ 

3 & 1.37(2) & 0.77(2) & $-$1.51(5) & $-$0.10(1) & 208(16)$^{{\mathrm c}}$ & 33(2)$^{{\mathrm c}}$ & 43(6)$^{{\mathrm c}}$ & 1198(121) & 202(8) & 1.14 & 1.05 & c & Strong silicate emission, \\ 
&&&&&&&&&&&&& fit too narrow \\

4 & 0.33(1) & 0.28(1) & $-$0.37(1) & $-$0.19(3) & 154(9) $^{{\mathrm c}}$& 45(4)$^{{\mathrm c}}$ & 32(5)$^{{\mathrm c}}$ & 0(0) & 102(5) & $-$1.03$^{{\mathrm f}}$ & $-$1.14$^{{\mathrm f}}$ & b & Poss.~13$\mu$m absorption band \\ 

5 & 0.10(1) & 0.08(1) & $-$0.57(3) & $-$0.22(3) & 208(8) & 57(4) & 29(4) & 0(0) & 59(4) & 0.06 & $-$0.09 & b & Long $\lambda$ fit probably wrong \\

6 & 6.16(14) & 5.31(13) & $-$0.33(1) & $-$0.12(2) & 137(5)$^{{\mathrm c}}$ & 38(2)$^{{\mathrm c}}$ & 24(4)$^{{\mathrm c}}$ & 0(0) & 103(4) & $-$0.11 & $-$0.18 & b & Poor 8--11$\mu$m fit \\

7 & 0.35(1) & 0.32(1) & $-$0.14(1) & $-$0.15(2) & 152(7)$^{{\mathrm c}}$ & 50(4)$^{{\mathrm c}}$ & 23(3)$^{{\mathrm c}}$ & 0(0) & 100(4) & $-$1.13$^{{\mathrm f}}$ & $-$1.10$^{{\mathrm f}}$ & b & Large discontinuity at join \\

8$^{{\mathrm e}}$ & 0.31(1) & 0.27(1) & $-$0.36(1) & $-$0.11(1) & 62(4) & 0(0) & 1(1) & 554(56) & 27(3) & $-$0.31 & $-$0.29 & b & \\

9 & 0.35(1) & 0.53(2) & 1.02(4) & $-$0.20(3) & 237(8) & 38(4) & 25(4) & 1119(114) & 144(6) & 0.99 & 1.33 & a & Dubious 6.8$\mu$m and 7.5$\mu$m fits \\

10 & 0.36(1) & 0.33(1) & 0.04(1) & $-$0.06(1) & 168(11)$^{{\mathrm c}}$ & 35(3)$^{{\mathrm c}}$ & 23(3)$^{{\mathrm c}}$ & 609(61) & 126(6) & $-$0.43 & $-$0.34 & b & Probably peak in nebulosity \\

11 & 0.36(2) & 1.89(10) & 4.34(31) & $-$0.62(8) & 984(45) & 350(67) & 73(11) & 1996(202) & 479(28) & 2.07 & 3.14 & a & Very deep absorption features \\

12$^{{\mathrm e}}$ & - & - & - & - & 414(76)$^{{\mathrm c}}$? & 0(0)$^{{\mathrm c}}$? & 19(5)$^{{\mathrm c}}$? & - & 143(16)? & - & - & - & Very dubious measurements: \\
&&&&&&&&&&&&& peaks on bad column\\

13 & 0.04(1) & 0.07(2) & 1.35(22) & $-$0.20(3) & 348(51)$^{{\mathrm c}}$ & 146(46)$^{{\mathrm c}}$ & 69(11)$^{{\mathrm c}}$ & 2496(291) & 185(7) & 0.55 & 1.09 & a & Good fit but very faint \\
\hline
Cha I 1$^{{\mathrm e}}$ & 0.71(2) & 0.65(2) & $-$0.18(1) & $-$0.07(1) & 0(0) & 0(0) & 7(1) & 548(55) & 0(0) & $-$0.32 & $-$0.28 & b & Poor fit, due to large \\
&&&&&&&&&&&&& discontinuity at join \\

2 & - & - & - & - & - & - & - & - & 150(6) & 1.81 & - & a & \\

3 & 0.14(1) & 0.13(1) & $-$0.37(2) & $-$0.07(1) & 144(10)$^{{\mathrm c}}$ & 36(3)$^{{\mathrm c}}$ & 10(2)$^{{\mathrm c}}$ & 0(0) & 0(0) & $-$0.40 & $-$0.52 & b & Noisy, poor fit at long $\lambda$ \\

4 & 0.04(1) & 0.02(1) & $-$1.88(11) & 0.13(2) & 225(25)$^{{\mathrm c}}$ & 87(9)$^{{\mathrm c}}$ & 0(0)$^{{\mathrm c}}$ & 0(0) & 0(0) & $-$0.65 & $-$1.09 & c & Very faint, low S/N \\ 

5 & 3.58(6) & 3.59(6) & 0.00(1) & $-$0.12(2) & 41(2)$^{{\mathrm c}}$ & 0(0)$^{{\mathrm c}}$ & 7(1)$^{{\mathrm c}}$ & 283(28) & 0(0) & $-$0.43 & $-$0.37 & b & Strong 11$\mu$m emission \\

6 & 3.18(6) & 1.93(4) & $-$1.32(5) & $-$0.04(1) & 0(0) & 0(0) & 0(0) & 0(0) & 0(0) & $-$0.85 & $-$1.07 & c & Silicate fit too narrow \\ 
\hline
\end{tabular}

\begin{list}{}{}
\item[$^{{\mathrm a}}$] Values of $\tau_{\mathrm {Si}}$ between $-$0.8 and 0.6 are considered to be questionable, due to the presence of significant silicate emission and absorption (see discussion in Sect.~\ref{sec:si_prof}).
\item[$^{{\mathrm b}}$] Classification scheme proposed in Sect.~\ref{sec:si_prof}.
\item[$^{{\mathrm c}}$] Short wavelength features fitted using linear
continuum, fitted from 5.6--7.9$\mu$m, in cases where the automatic fitting
procedure produced errors (see Sect.~\ref{sec:badfits}).
\item[$^{{\mathrm d}}$] Feature fitted manually as spectrum was incomplete.
\item[$^{{\mathrm e}}$] Source fell on bad column on array: data from these sources is not considered to be especially accurate.
\item[$^{{\mathrm f}}$] No K-band data: $\alpha$ evaluated using $\lambda_1 = 8.0$$\mu$m.
\end{list}

\end{table}
\end{landscape}


\begin{figure}
	\resizebox{\hsize}{!}{
	\begin{turn}{270}
	\includegraphics{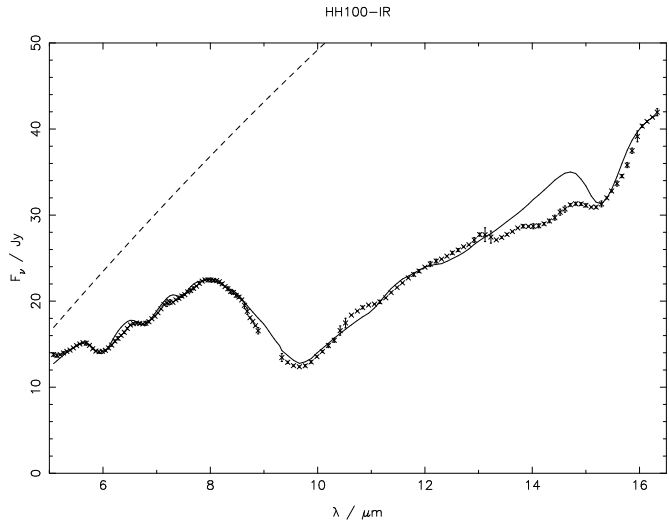}
	\end{turn}
	}
	\caption{The spectrum of \object{HH100-IR}, with the fitted spectrum (solid line) and continuum (dashed line).  Note the poor fit to the CO$_2$ feature at 15$\mu$m as a further example of why the CO$_2$ feature was fitted independently.}
	\label{fig:hh100ir}
\end{figure}

\noindent detected, this was found to be the most reliable and consistent method of spectral fitting.

In order to check the instrumental calibration and the validity of this fitting method, comparisons were made with previous observations of the bright, well-studied object \object{HH100-IR} (RCrA 5, see Fig.\ref{fig:hh100ir}).  Whittet et al.~(\cite{whittet96}) find a silicate optical depth of 1.21$\pm$0.05, which is somewhat less than the value of 1.35$\pm$0.05 measured here: this is almost certainly due to the fact that the iterative fitting procedure used here allows the continuum to drift away from the observed spectrum slightly.  As discussed in Sect.\ref{sec:fitting}, this is due to absorption across the entire wavelength range and is a benefit made possible by the broad wavelength coverage of the CVF.  Whittet et al.~(\cite{whittet96}) also find a H$_2$O ice column density of 2.4$\times$10$^{18}$cm$^{-2}$, based on observations of the 3$\mu$m stretching mode.  Adopting a band strength of $A$=1.2$\times$10$^{-17}$cm molecule$^{-1}$ (Gerakines et al.~\cite{gerakines95}), we evaluate the column density $N$ as:
\begin{equation}
N = \frac{\int \tau(\lambda) {\mathrm d}\lambda}{\lambda_{\mathrm {peak}}^2 A}
\end{equation}
where $\int \tau(\lambda) {\mathrm d}\lambda$ is the equivalent width.  This gives a value of 4.8$\times$10$^{18}$cm$^{-2}$, approximately double the value obtained from the stretching mode.  This is consistent with the ``6$\mu$m to 3$\mu$m paradox'' (Gibb et al.~\cite{gibb00}; Dartois \& d'Hendecourt \cite{dd01}), which commonly results in observations of these two bands producing column densities which differ by a factor of $\sim$2.  The recent work of Gibb \& Whittet (\cite{gw02}) suggests that the excess depth of the 6$\mu$m feature is the result of blending with a feature due to organic refractory matter.  They also find a strong correlation between the excess absorption in the 6$\mu$m water ice feature and that of the 4.62$\mu$m ``XCN'' feature; unfortunately the wavelength range of the CVF prevented observation of the ``XCN'' feature here.  Similarly, from observations of the 4.27$\mu$m stretching mode Nummelin et al.~(\cite{nummelin01}) infer a CO$_2$ ice column density of 6.2$\pm$0.6$\times$10$^{17}$cm$^{-2}$, which is broadly consistent with the value of 5.1$\times$10$^{17}$cm$^{-2}$ measured here (assuming a band strength of $A$=1.1$\times$10$^{-17}$cm molecule$^{-1}$, Gerakines et al.~\cite{gerakines95}).  Keane et al.~(\cite{keane01}) measure optical depths of $\tau_{6.0}$=0.23 and $\tau_{6.8}$=0.09, which are consistent with our results of 0.23$\pm$0.01 and 0.13$\pm$0.01 respectively.  It should be noted, however, that our value of the 8$\mu$m flux density, 1.05$\pm$0.02$\times$10$^{-16}$W cm$^{-2}$ $\mu$m$^{-1}$, is somewhat less than the value of $\sim$1.7$\times$10$^{-16}$W cm$^{-2}$ $\mu$m$^{-1}$ measured by Whittet et al.~(\cite{whittet96}).  This may indicate a possible error in the absolute flux calibration of the CVF data.  However, the measurement of spectral features depends only on the relative flux calibration from pixel to pixel, which is considered to be accurate throughout.

\subsubsection{Water Ice: Bending and Libration Modes}\label{sec:libration}
The observed equivalent widths of the 6$\mu$m bending and 13$\mu$m libration modes of water ice showed no significant correlation at all.  Further, the libration mode was not observed to correlate with any of the measured features.  Whilst there is evidence for this in the literature (eg.~Bowey at al.~\cite{baw98}) the most likely explanation is that the libration mode is poorly fitted, due to the wide variation in possible profiles.  The peak of this band has previously been found at $\sim$11$\mu$m in crystalline water, $\sim$12.5$\mu$m in amorphous water and at even longer wavelengths in mixtures with other molecules (Hagen et al.~\cite{htg83}; d'Hendecourt \& Allamandola \cite{dha86}; Cox \cite{cox89}).  Here a single profile was used (that of amorphous ice) and this probably resulted in poor fitting.  Unfortunately this section of the spectrum is strongly blended with the silicate feature, so this problem cannot be remedied without making further ad hoc assumptions about the silicate profile(s).

\subsubsection{The Unidentified Feature at 6.8$\mu$m}\label{sec:6.8um}
\begin{figure}
	\resizebox{\hsize}{!}{
	\includegraphics{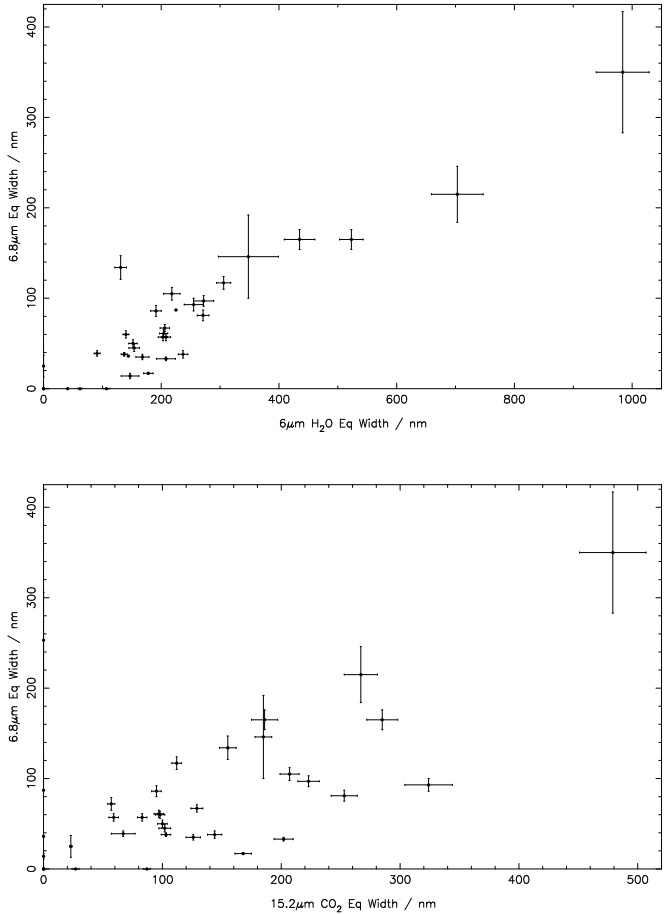}
	}
	\caption{The strength of the unidentified feature plotted against the bending modes of both water (top) and CO$_2$ (bottom) ices.}
	\label{fig:uir}
\end{figure}
As seen in Fig.\ref{fig:uir}, the strength of the unidentified 6.8$\mu$m feature is observed to correlate far more strongly with the neighbouring 6$\mu$m band of water ice than with the 15.2$\mu$m band of CO$_2$ ice.  This is consistent with previous observations, where the 6 and 6.8$\mu$m features usually correlate well (see Schutte et al.~\cite{schutte96} and the review by Schutte \cite{schutte97}).  This suggests that the carrier of the unidentified feature is probably a strongly polar ice, as its presence matches far more closely the strongly polar H$_2$O ice than the markedly less polar CO$_2$.  The more recent work of Keane et al.~(\cite{keane01}) found the peak position of this feature at different wavelengths towards different sources, and proposed that it consists of two (related) components.  Further detailed study of the 6.8$\mu$m profiles could yield a great deal more information about their chemistry.  However, at the low spectral resolution of the CVF the 6.8$\mu$m profiles are essentially constant across all the sources, and so such an investigation was not possible here.  One individual source is also worthy of note: GY262 ($\rho$ Oph E 3) shows a deep 6.8$\mu$m feature with no corresponding 6.0$\mu$m feature (the ``rogue'' point to the upper left in Fig.\ref{fig:uir}).

\subsubsection{11.2$\mu$m Feature}\label{sec:11um}
The depth of the measured 11.2$\mu$m feature was found to correlate strongly with the depth of the silicate feature, with a direct (negative) proportionality providing a good fit to the data.  Consequently the 11.2$\mu$m feature was interpreted to be an emissive shoulder on the silicate feature, rather than an independent feature due to another species: the presence of this feature narrows the silicate absorption profile slightly.  The silicate profile has previously been found to vary depending on the composition and structure of the silicate grains (eg.~Demyk et al.~\cite{demyk00}), so such a shoulder is not unexpected.  However, 3 of the sources (RCrA 1, Ser A 6 and Cha I 3) show a significant 11.2$\mu$m emission feature without any significant silicate absorption feature: these features may be attributable to emission from crystalline silicates (Bregman et al.~\cite{bregman87}; Campins \& Ryan \cite{cr89}).

\subsubsection{CO$_2$ Ice Profiles}\label{sec:co2prof}
As mentioned in Sect.~\ref{sec:fitting}, the CO$_2$ ice profile is a very sensitive diagnostic of the ice environment.  Whilst the spectral resolution of the CVF was not sufficient to study the profiles in great detail, it is worth noting that a significant long wavelength wing was present in almost all the observed absorption profiles.  This would seem to indicate that a large fraction of the CO$_2$ ice observed exists in a polar (H$_2$O-rich) phase, which is characterised by this long wavelength wing (Gerakines et al.~\cite{gerakines99}).  However quantifying the relative abundances of the different phases was not possible at this low spectral resolution.

\subsubsection{Spectral Classification and Silicate Profiles}\label{sec:si_prof}
\begin{figure}
	\resizebox{\hsize}{!}{
	\begin{turn}{270}
	\includegraphics{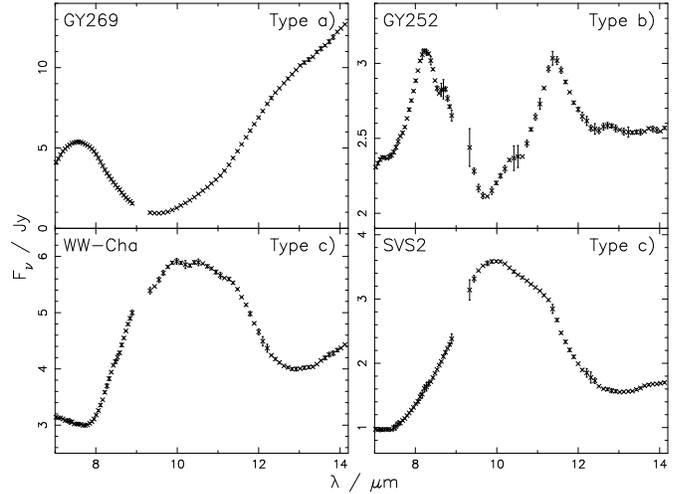}
	\end{turn}
	}
	\caption{Examples of observed silicate profiles: emission profiles are generally broader than those seen in absorption and composite emission/absorption profiles are also seen.  ``Type'' refers to the classification scheme described in Sect.~\ref{sec:si_prof}.}
	\label{fig:sil}
\end{figure}
If we look at the shapes of the spectra, it appears natural to divide the sources into 3 distinct groups based on the strengths of the spectral features observed.  An example of each type is shown in Fig.\ref{fig:sil} and the classifications are included in Table \ref{tab:res1}.
\begin{list}{}{}
\item[{\bf a)}] The first group show deep absorption features due to all of the ices and also due to the silicates (typically these have $\tau_{\mathrm {Si}} \gtrsim 0.6$).  These objects are interpreted to be heavily embedded objects, showing strong absorption due to cold foreground material.  
\item[{\bf b)}] The second group shows weaker ice absorption features and weak silicate features ($-0.8\lesssim\tau_{\mathrm {Si}}\lesssim0.6$) which are, in general, not especially well fitted.  These are interpreted as less heavily embedded objects, showing less foreground absorption.  The poor silicate fits are probably due to complications caused by combined silicate emission and absorption.  
\item[{\bf c)}] The third group show strong silicate emission ($\tau_{\mathrm {Si}}\lesssim-1.0$) and little or no ice absorption.  It should be noted that in the 2 good S/N spectra showing strong silicate emission, the observed emission profiles were significantly broader than both the observed absorption profiles and the profile used to fit the data (see Fig.\ref{fig:sil}).  (However it should also be noted that one of these 2 sources, \object{SVS2}, is an unresolved double.)  These are probably young stars which have shed most of their circumstellar matter, retaining only a little hot, optically thin silicate dust around them (possibly in a disc).  It therefore seems plausible that these three groups might represent an approximate evolutionary sequence, with the YSOs shedding circumstellar material as they evolve.
\end{list}

\begin{figure}
	\resizebox{\hsize}{!}{
	\begin{turn}{270}
	\includegraphics{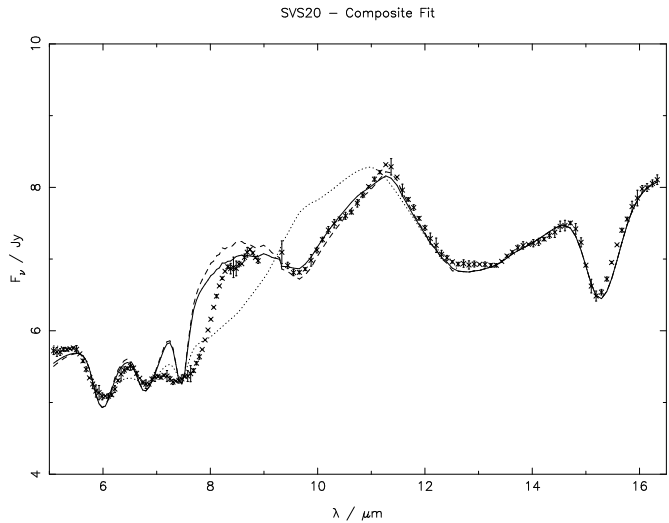}
	\end{turn}
	}
	\caption{Composite silicate profiles fitted to the spectrum of \object{SVS20}.  The solid line shows a composite fit with $\tau_{\mathrm {abs}}$=0.59 and $\tau_{\mathrm {em}}$=$-$0.33, the dashed line a composite fit with $\tau_{\mathrm {abs}}$=0.88 and $\tau_{\mathrm {em}}$=$-$0.48 (and a different continuum).  The single profile fit reported in the results table, with $\tau_\mathrm{Si}$=$-$0.33, is shown as the dotted line.}
	\label{fig:SVS20}
\end{figure}

In order to test this the spectrum of \object{SVS20} (Ser B 6) was re-fitted using a composite silicate profile, consisting of the standard absorption profile used above superimposed on the broader emission profile of \object{SVS2} (Ser B 3 - see Fig.\ref{fig:sil}).  As seen in Fig.\ref{fig:SVS20} this provided a good fit to the observed silicate profile, indicating that both silicate emission and absorption probably do occur, with different profiles seen in absorption and emission.  The physical interpretation of this is that these objects have a hot region of optically thin silicates around them but remain embedded in the cloud, which results in foreground absorption due to silicates and ices being superimposed on the ``intrinsic'' silicate emission feature in the spectra.  

It should be noted that while it is possible to fit such profiles to these sources, they are not well constrained.   As can be seen by comparing the 2 composite profiles in Fig.\ref{fig:SVS20} there is a degeneracy between the strength of this ``flat-topped'' silicate profile and the continuum strength, and a great deal of information is assumed about the silicate profiles themselves.  Consequently this investigation was not pursued for all the sources.  Where $-0.8\lesssim\tau_{{\mathrm Si}}\lesssim0.6$ the values in Table \ref{tab:res1} represent the correct optical depth at 9.7$\mu$m, but are usually in error towards the wings of the silicate feature.  Such values are useful in studies of general trends, but they do not accurately describe the entire silicate feature: a single number cannot fully describe the complicated nature of these features.

This also provides more direct evidence, similar to that of Whittet et al.~(\cite{whittet88}), that the extensive scatter in measurements of the $A_V/\tau_{\mathrm{Si}}$ ratio towards YSOs may be caused by complications due to silicate emission.  Direct measurements of the silicate optical depth will significantly under-estimate the depth of foreground material in cases (such as SVS20) where optically thin silicate emission is also present.  Rieke \& Lebofsky (\cite{rl85}) derive $A_V/\tau_\mathrm{Si} \simeq 17$, whilst noting the problem posed by ``intrinsic'' silicate emission, and such emission does indeed lead to errors when evaluating $A_V$ from the silicate depth.  As an example, the ISOCAM survey of Bontemps et al.~(\cite{bontemps01}) derives values of $A_V \simeq 28$  for \object{GY252} ($\rho$ Oph E 1) and $\simeq 24$ for \object{GY262} ($\rho$ Oph E 3); the X-ray survey of Imanishi et al.~(\cite{ikt01}) derives slightly lower values of $\sim24$ and $\sim18$ respectively.  However \object{GY252}, as seen in Fig.\ref{fig:sil}, clearly shows a ``composite'' silicate profile whereas GY262 does not, and the respective silicate optical depths of 0.52 and 1.84 are clearly inconsistent with a single $A_V/\tau_\mathrm{Si}$ ratio.  Unless the emitting and absorbing components of such silicate features can be separated unambiguously obtaining a single $A_V/\tau_\mathrm{Si}$ ratio will not be possible.

\subsubsection{Spectral Index}\label{sec:class}
Of the 41 objects to which spectral indices were assigned 20 were found to be class I objects ($\alpha_{\mathrm {cont}}>0$) and 17 to be class II ($-1.5<\alpha_{\mathrm {cont}}<0$).  In addition to this 3 objects ($\rho$ Oph E 1 and 3, Ser B 2) were found to lie very close to the class I/II border (so-called transition objects: Andr\'{e} \& Montmerle \cite{am94}; Greene et al.~\cite{greene94}).  The lack of any class III objects is a selection effect, but it seems to indicate that all of the observed sources are within the clouds, as field stars would show little or no infrared excess and would appear as class III objects in this survey.  The distribution of class I and II objects is approximately constant throughout all four regions, and no significant trends involving spectral class were found.  Evaluation of mid-IR spectral indices appears to support the finding of Bontemps et al.~(\cite{bontemps01}) that mid-IR spectral indices do not discriminate strongly between class I and class II objects, and a classification scheme similar to that in Sect.~\ref{sec:si_prof} seems preferable when considering spectroscopic data.

Comparing the index-based spectral classes to the spectral feature grouping above, we see that the majority of the class I sources belong to group a) and the majority of the class II sources to group b).  However it is notable that a significant number (6) of the 23 group a) objects are either class II or transition objects.  This in keeping with the convention that class I objects are the youngest (Lada \cite{lada87}), but it may be significant that there are two distinct types of spectrum which both result in class II SEDs.  Objects showing strong foreground absorption and objects showing little foreground absorption can both present class II continuum SEDs.  This could be due to geometry of the YSOs relative to the absorbing clouds, or could be an effect intrinsic to the sources: spectroscopy along a pencil-beam cannot distinguish these.  Whilst a larger sample size would constrain this problem further, it does seem that mid-IR observations alone do not provide strong constraints on the ``true'' spectral class of YSOs.


\section{Variations Between Target Regions}\label{sec:regvar}
In general, the most striking regional variation is between the sources in Cha I and those in the other three regions.  All of the sources in Cha I, except for \object{ISO-ChaI 192} (Cha I 2), show little or no absorption due to volatile ices and show silicates either in emission or an emission/absorption composite.  As it lay outside the CVF1 images, due to rotation of the spacecraft between the two sets of observations, we only have a 9.3--16.3$\mu$m spectrum for ISO-ChaI 192.  However it shows CO$_2$ ice absorption and also appears to show deep silicate absorption ($\tau_\mathrm{Si} \sim 2$--4), so it is probably an embedded object.  Persi et al.~(\cite{persi99}) found that \object{ISO-ChaI 192} lies in a dense core and that it is probably the engine for the bipolar outflow detected by Mattila et al.~(\cite{mlt89}).  Jones at al.~(\cite{jones85}) found that several other YSOs have formed around the edge of this core, rather than at its centre: these were apparently formed during a burst of star formation triggered by a wind from the nearby star \object{HD97300}.  Our results are consistent with this: \object{ISO-ChaI 192} appears to be heavily embedded, and has a very red SED.  The rest of the surrounding objects show similar spectral characteristics, and appear to be surrounded by far less circumstellar/foreground material than \object{ISO-ChaI 192}.

12 of the 20 objects in Serpens show deep absorption features, 7 show composite silicate profiles, and one (\object{SVS2}) shows silicate emission.  It is clear that Serpens contains a wide range of YSOs, ranging from the most heavily embedded object observed (Ser B 11) to a source showing silicate emission.  By contrast almost all of the sources in both RCrA and $\rho$ Oph show deep absorption features, with only 1 of the 7 sources in RCrA and only 2 of the 10 sources in $\rho$ Oph showing composite silicate profiles.  Given that YSOs are generally expected to ``sweep out'' their circumstellar material as they evolve (Shu et al.~\cite{shu87}), this seems to imply that the observed star formation is at a similar evolutionary stage in both RCrA and $\rho$ Oph, probably an earlier stage than that in Serpens.  However, given the broad variation in the sources observed in Serpens, this conclusion remains somewhat tentative.   It should also be noted that the observed fields (of 2$\arcmin$$\times$2$\arcmin$) are much smaller than the star-forming clouds, which typically extend over several square degrees of the sky, so the populations observed here may not be representative of the entire clouds.

\subsection{Silicate Depth and H$_2$O Bending Mode}\label{sec:tau_water}
\begin{figure}
	\resizebox{\hsize}{!}{
	\begin{turn}{270}
	\includegraphics{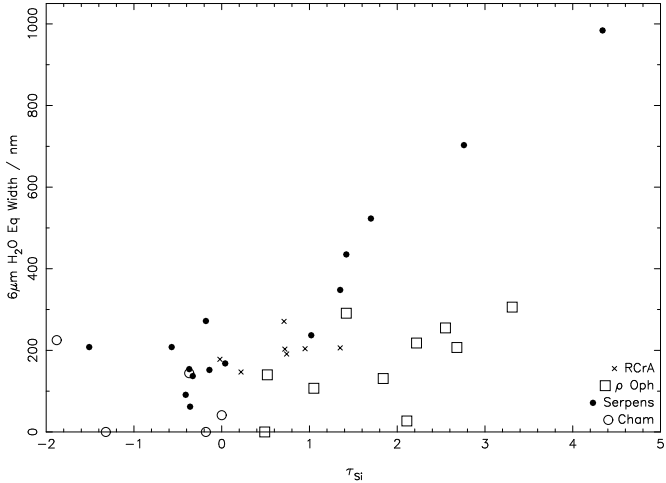}
	\end{turn}
	}
	\caption{The strength of the bending mode of water ice plotted against the silicate optical depth.  Different regions are distinguished by different symbols, as shown in the legend.}
	\label{fig:silh2o}
\end{figure}
Figure \ref{fig:silh2o} shows the measured equivalent widths of the 6$\mu$m water ice feature plotted against the optical depth of the silicate feature: clear regional variations are apparent.  As discussed above, all but one of the sources in Cha I are characterised by little or no ice absorption and by silicate emission; the other 3 regions show absorption from both silicates and ices with only a few examples of weak silicate emission.  The sources in RCrA are characterised by similar absorption strengths throughout, whereas far deeper absorption and variation is observed in both $\rho$ Oph and Serpens.  It is clear from Fig.\ref{fig:silh2o}, however, that the sources in $\rho$ Oph show markedly less water ice absorption relative to silicates than those in Serpens.  This could be due to sublimation effects, indicating a difference in the thermal structure and history of the two clouds, or could represent a real difference in the chemical composition of the two clouds; without measuring the corresponding gas-phase lines we cannot say which.  Sub-millimetre observations of gas-phase H$_2$O (Ashby et al.~\cite{ashby00}) do show an under-abundance of gaseous H$_2$O in $\rho$ Oph A relative to S140, but to the best of our knowledge no similar direct comparisons between the sources observed here exist in the literature.  It is also worth noting that the silicate:water ice ratios appear to be approximately constant for all the different embedded sources in both $\rho$ Oph and Serpens.  This indicates that in both clouds the composition of the absorbing intracloud material is roughly constant along different lines of sight, a result consistent with the previous Serpens work of Eiroa \& Hodapp (\cite{eh89}).  

\subsection{H$_2$O and CO$_2$ Bending Modes}\label{sec:ices}
\begin{figure}
	\resizebox{\hsize}{!}{
	\begin{turn}{270}
	\includegraphics{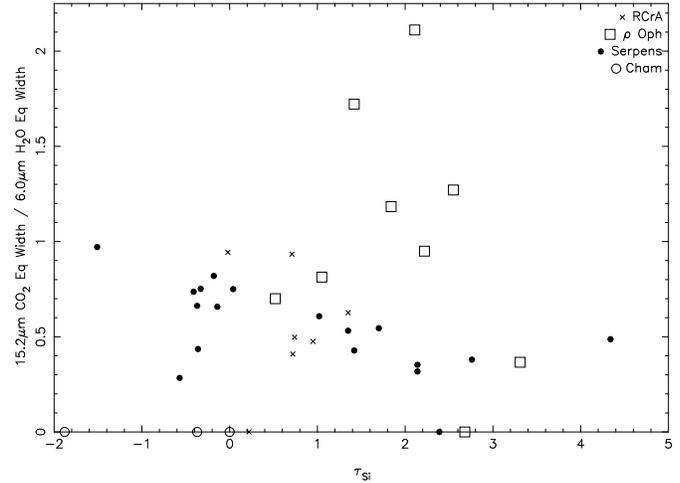}
	\end{turn}
	}
	\caption{The ratio of the CO$_2$:H$_2$O ice equivalent widths plotted against silicate optical depth.  This ratio is directly proportional to the ratio of ice column densities: adopting the band strengths from Gerakines et al.~(\cite{gerakines95}) results in an equivalent width ratio of 1 corresponding to a column density ratio of $\simeq$0.17.}
	\label{fig:615}
\end{figure}

Figure \ref{fig:615} shows the ratio of the CO$_2$:H$_2$O ice equivalent widths (which is proportional to the corresponding ratio of column densities) plotted against the silicate optical depth.  The results from Serpens and RCrA are fairly well correlated, with CO$_2$:H$_2$O ice column density ratios ranging from 0 to 0.16 (adopting the band strengths from Gerakines et al.~\cite{gerakines95}).  There is a possible trend showing the CO$_2$:H$_2$O increasing with decreasing silicate optical depth, but this is somewhat unclear in these data.

The data from $\rho$ Oph, however, show a far greater scatter with CO$_2$:H$_2$O ice column density ratios ranging from 0 to 0.4.  This is consistent with previous observations: for example Chiar et al.~(\cite{chiar94}, \cite{chiar95}) found a greater scatter in column densities of CO and H$_2$O ices against $A_V$ in $\rho$ Oph than in Serpens, RCrA or Taurus.  The reasons for this are unclear but it seems likely that, as suggested by Chiar et al.~(\cite{chiar95}), local conditions around the young stars in $\rho$ Oph play a role.  Detailed study of the gas-phase abundances of CO$_2$ and H$_2$O towards these objects will further constrain this problem, but without such observations or further knowledge of the envelope structures we merely note the presence of this discrepancy here.


\section{Conclusions}\label{sec:conc}
ISO-CVF spectroscopy of four low-mass star-formation regions has been performed, and a total of 42 different sources were observed.  The resulting low-resolution 5--16.5$\mu$m spectra have been analysed, and a number of different conclusions drawn:
\begin{itemize}
\item The profile of the 9.7$\mu$m silicate absorption feature varies somewhat from source to source, but usually shows a weak emissive shoulder at $\simeq$11.2$\mu$m.  Significantly broader profiles are observed in emission.  A number of sources appear to show ``composite'' profiles, with absorbing and emitting components superimposed.
\item The presence of a long wavelength wing on the 15.2$\mu$m CO$_2$ ice band in most of the sources suggests that a significant fraction of the CO$_2$ ice observed exists in a polar (H$_2$O-rich) phase.
\item The strength of the unidentified 6.8$\mu$m band is observed to correlate far more strongly with the 6.0$\mu$m water ice band than with the 15.2$\mu$m CO$_2$ ice band.  This suggests, in a manner consistent with previous observations, that the carrier of this band is a strongly polar ice.
\item Compared to the other 3 regions (which show similar spectral characteristics) the sources observed in Cha I are somewhat anomalous, showing comparatively little foreground absorption.  However this is consistent with previous studies, which indicate that the observed Cha I sources represent an unusual cluster of YSOs which have formed at the edge of a dense cloud core.
\item The $\rho$ Oph cloud appears to have a systematic under-abundance of water ice, relative to silicates, relative to both RCrA and Serpens.
\item The CO$_2$:H$_2$O ice ratios towards the sources in $\rho$ Oph show considerably greater scatter than in the other regions, with values ranging from 0 to 0.4.  This may be due to local conditions around the YSOs in $\rho$ Oph but more data is needed to constrain this problem further.
\end{itemize}
The most important limiting factor in this study is obviously the low spectral resolution of the CVF, which essentially prevents any detailed study of absorption profiles, particularly around the CO$_2$ ice feature.  However the use of the CVF enabled the study of a large number of objects, many of which would not have been observable at higher spectral resolution due to the low S/N.  Consequently we have obtained a unique data set, with a large sample size and broad wavelength coverage.  This enabled simultaneous study of the dominant mid-IR spectral features along the line-of-sight to a large number of YSOs, in a manner not possible using other spectroscopic methods.
Further detailed study of the ice absorption profiles will undoubtedly yield more information regarding the chemistry around these objects, and study of the silicate profiles may yield more information about the dust composition and the structure of the dust envelopes.  


\begin{acknowledgements}
We thank Tom Greene for his work in planning the observations and for helpful comments.
We thank Andy Longmore for useful comments about the manuscript 
and also thank Sylvain Bontemps for useful discussions.
Part of this work formed part of the MPhys dissertation of RDA at the University of Edinburgh, and RDA also thanks the UKATC for funding a vacation studentship.
Finally, we thank an anonymous referee for helpful advice which greatly improved the clarity of the paper.
\end{acknowledgements}


\appendix

\section{Atlas of Spectra}\label{sec:app}
All of the spectra are presented in this appendix.  In each plot the data points are shown as crosses, with the fits as solid lines and the fitted continua as broken lines.  The spectra are arranged in the same order as in Table \ref{tab:sources}.  The error bars shown are the statistical errors on the data points and do not include the absolute flux calibration errors discussed in Sect.~\ref{sec:fitres}.  Please also note that the fits shown are the ``complete'' fits, covering the complete wavelength range: in some cases (as indicated in Table \ref{tab:res1} and discussed in Sect.~\ref{sec:badfits}) the narrower features were fitted independently and these fits are not shown here.

\end{document}